\documentclass[fleqn,twoside]{article}
\usepackage{espcrc2}
\usepackage{graphicx}
\usepackage{graphicx}
\usepackage{amssymb}

\title{Transition Radiation Spectroscopy with Prototypes of the ALICE TRD}
\author{O.~Busch\thanks{\mbox{O.~Busch, GSI, Planckstr. 1, 64291 Darmstadt, Germany}
\mbox{\textit{E-mail address:} o.busch@gsi.de}} 
for the ALICE TRD collaboration\thanks{A list of the members of the ALICE TRD collaboration is 
given at the end of this paper}}

\begin{document}

\begin{abstract}
We present measurements of the transition radiation (TR) spectrum produced in an irregular 
radiator at different electron momenta. The data are compared to simulations of TR from a regular radiator. 
\end{abstract}

\maketitle

\section{Motivation}

Measurements of TR yield and spectra for a large variety of regular radiator configurations 
have been presented by different authors. In general, data confirm the theoretical 
predictions \cite{FabjanStruc}, although total TR yield and dependence on the momentum of the emitting particle 
are not always reproduced \cite{CherrySpectrometer}\cite{CherrySilicon}. 
The ALICE Transition Radiation Detector (TRD) consists of 
540 drift chambers with mixed fibre/foam radiators. The expected performance of the 
detector in heavy-ion collisions is investigated in detailed 
simulations of the detector response. An accurate numerical treatment of TR production in an 
irregular-layered radiator requires knowledge of the distributions of material thickness and 
spacing in the medium \cite{GaribianIrreg}, information usually 
not easy to quantify with precision. A realistic implementation of the ALICE TRD radiator in the simulations 
requires measurements of the spectral distribution of the emitted TR. 

\section{Setup and simulation}

\begin{figure}[hbt]
\includegraphics[width=.48\textwidth, height=.1\textheight]
{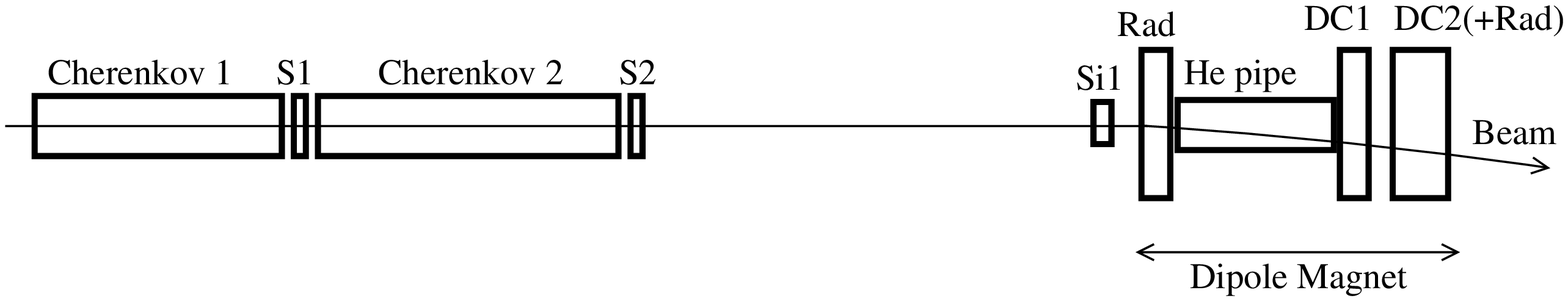}
\vspace{-1.0cm}
\caption{Schematic view of the setup (not to scale). To dissociate TR photons and beam particles, 
radiator and DC1, DC2 are separated and placed in a dipole magnet.}
\label{setup}
\end{figure}

The measurements were carried out using a secondary beam at the CERN PS. A schematic drawing of the setup is 
shown in Fig.~\ref{setup}. Two threshold Cherenkov detectors provide offline electron-pion discrimination. 
We use two prototype drift chambers (DC) described in \cite{AntonProc}, 
operated with the standard gas mixture for the TRD, Xe,CO$_2$(15\%), and read out via a low-noise 
fast preamplifier-shaper and FADC. The radiator is composed of 8 pure polypropylene fibre mats, 
corresponding to 3.6~cm total thickness, in a box of 6~mm carbon fibre-enforced Rohacell$^{\copyright}$ HF71.
It is separated from DC1 by a He-filled plexiglas tube, 80~cm long, with 2 aluminum coated 
mylar foils of 10~$\mu$m thickness serving as gas barriers. To deflect and separate the beam from the TR photons, 
radiator, He~pipe and the DCs are placed in a dipole magnet. For beam momenta of 1.5, 2, and 3~GeV/c 
the magnetic field strength is B=0.42, 0.42, and 0.56~T, respectively. In addition, runs at B=0 are carried 
out for each momentum. \par
TR production is simulated tuning a regular foil stack configuration 
to reproduce the measurements. The doubly differential TR yield 
$\frac{\mathrm{d}^2 W}{\hbar \mathrm{d} \omega \mathrm{d}{\Omega}}$ \cite{FabjanStruc} is integrated numerically 
over the solid angle $\Omega$. The parameters (220 polypropylene foils of 12~$\mu$m thickness, 
separated by air gaps of 100~$\mu$m) reflect typical dimensions of the radiator materials \cite{TDR}, but 
are not unambiguously determined. We use tabulated X-ray cross sections from \cite{XSec} to 
calculate photon absorption in the materials and the chamber gas. 

\section{Charge reconstruction} \label{sec:Q}

The signal induced by beam particles and TR photons is measured on a row of 
8 readout pads. The integrated pulse height is a measure of the deposited charge. 
Photon scattering, the angular spread of the beam 
and the Lorentz angle of the ionization drifting in the detector result in 
a wide distribution of the charge from absorbed TR and beam ionization over the pads. 
For each incident electron a TR cluster search is performed, connecting time 
intervals with signal over threshold on adjacent pads. Local minima in the pulse height 
distribution occuring simultaneously on adjacent pads are detected to resolve multiple overlapping photons. The position of the incident beam is identified 
from the signal in DC2. To avoid contamination of the measured spectra due to overlap of 
beam and TR, a separation of 2 pads or more is required. In case the distance is exactly 2 pads, the 
signal measured on the interjacent pad can not be unambiguously assigned to the beam or the TR cluster. 
In this case the cluster is rejected unless the contribution of the interjacent pad to the 
total charge of the cluster is less than 5\%.

\begin{figure}[hbt]
\includegraphics[width=.45\textwidth]
{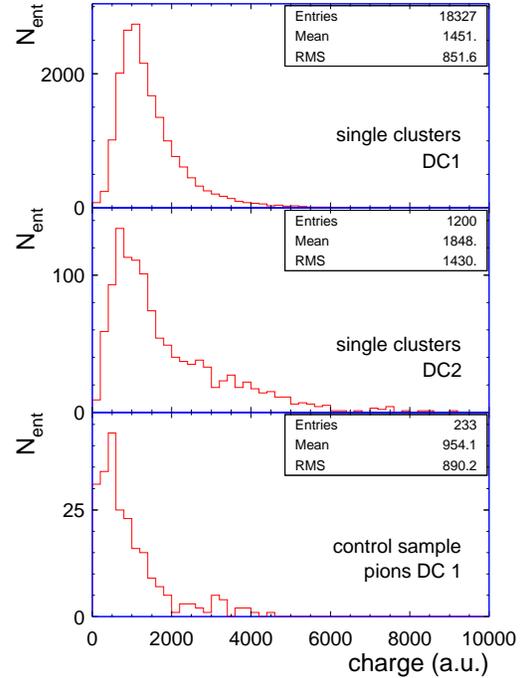} 
\vspace{-1.0cm} 
\caption{TR charge spectra for 2 GeV/c. Upper and center panel: single clusters in DC1 and DC2. 
Lower panel: charge of accepted 'TR' clusters for incident pions.} \vspace{-1.0cm}
\label{qplots} 
\end{figure} 

In Fig.~\ref{qplots} we present the TR charge spectra measured 
in DC1 and DC2 (upper and center panel). A smaller number of photons with higher average energy 
is detected in DC2, since most of the TR is deposited in DC1 or absorbed 
in the material before DC2, and only a fraction of hard photons 
penetrates into DC2. To assess the noise rejection power of the TR 
search algorithm we apply it to the sample of incident pions. The resulting charge spectrum 
of fake TR clusters is shown in the lower panel of Fig.~\ref{qplots}. 
Comparing the number of entries to the total size of the pion sample, the probability 
to produce a fake TR is found to be smaller than 2\%.

\section{Cluster number distribution}

\begin{figure}[hbt]
\begin{tabular}{c} 
\begin{minipage}{.43\textwidth}
\vspace{-.1cm}\includegraphics[width=1.0\textwidth,bbllx=0,bblly=0,
bburx=567,bbury=365,clip=]{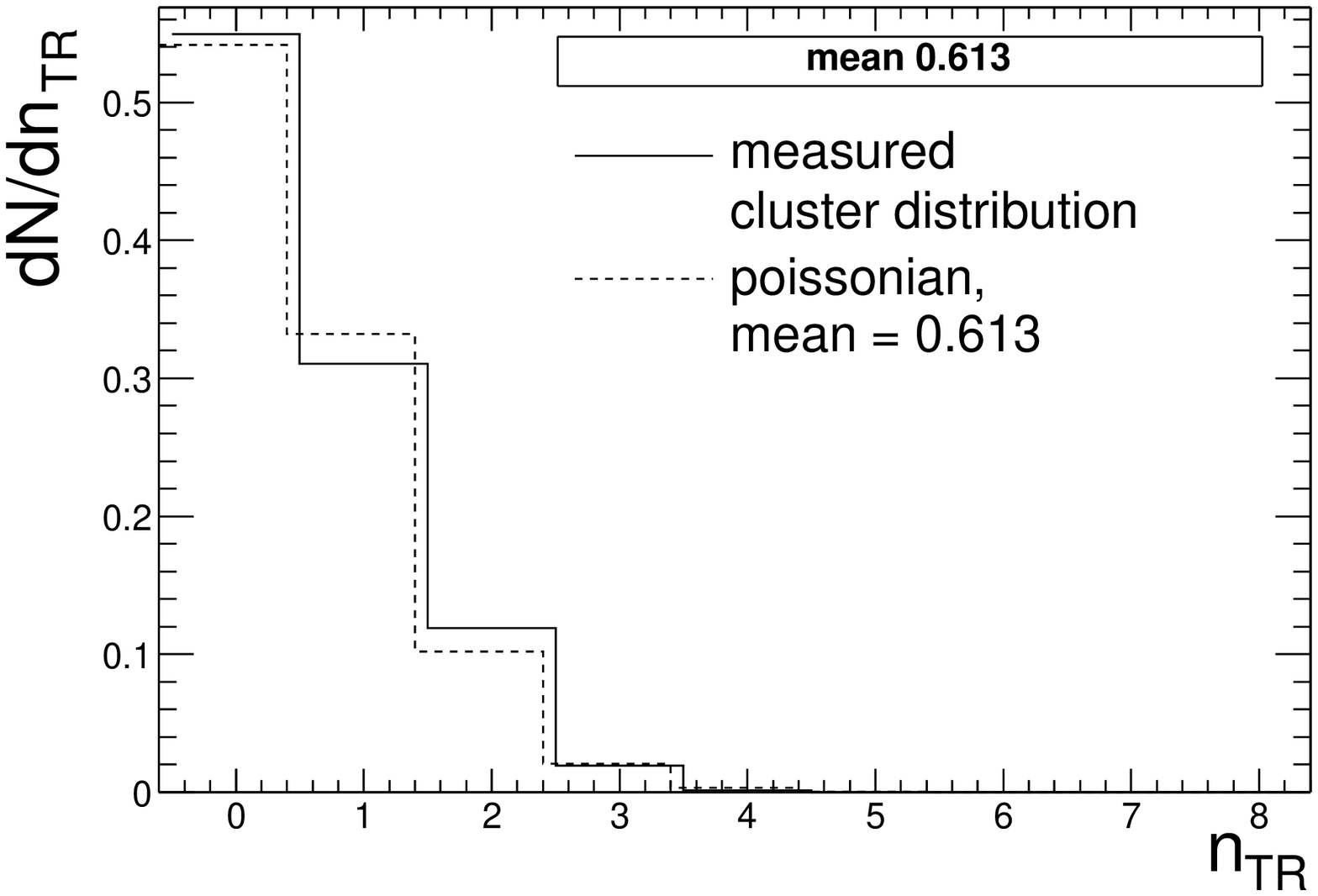} 
\end{minipage} \\
\begin{minipage}{.43\textwidth}
\includegraphics[width=1.0\textwidth, bbllx=0,bblly=0,
bburx=567,bbury=365, clip=]{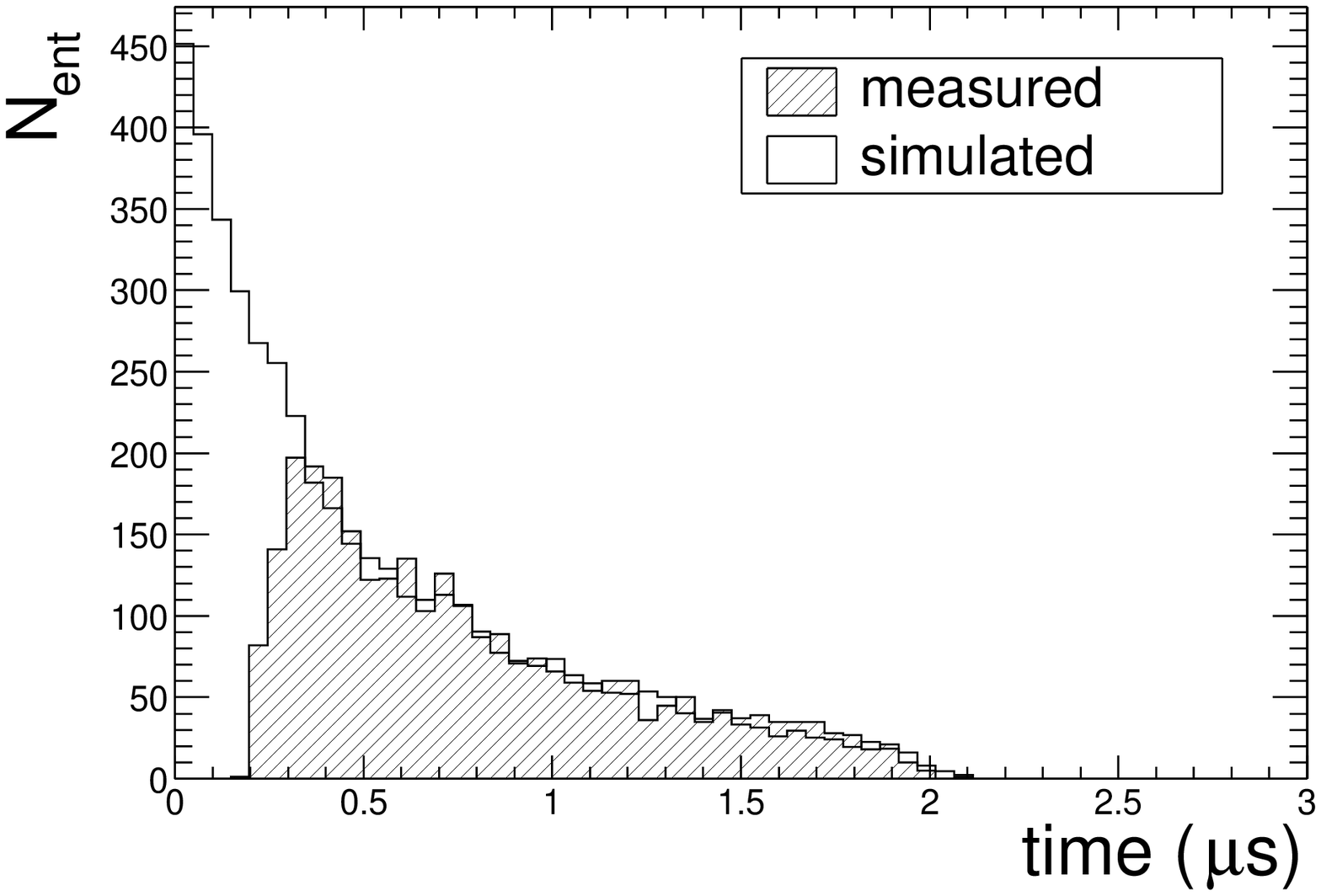}  
\end{minipage}
\end{tabular}
\vspace{-0.8cm}
\caption{Upper panel: photon number distribution for 2~GeV/c beam momentum, compared to a Poissonian distribution with 
equal mean. Lower panel: cluster overlap. Measured time interval between 2 TR photons for 2~GeV/c beam momentum 
compared to the distance between 2 TR photons in simulations.}
\label{dist2photons}
\end{figure}

In the upper panel of Fig.~\ref{dist2photons} we present the normalised distribution of the detected photon 
number per incident electron for 2~GeV/c beam momentum. The shape 
of the distribution compares well with a Poissonian, indicated by the 
dashed curve. On average 0.61 photons per incident 
electron are detected. This number is smaller than expected from 
simulations ($\sim$0.8 for the measured momenta). To some extent, TR overlap due to 
the finite time response of the detector and associated 
electronics accounts for this discrepancy, as illustrated in the lower panel. 
The minimum time interval between two TR photons resolved in the 
measurements is 0.2~$\mu$s. Comparing to the distribution obtained in simulations with 
ideal 2-cluster resolution we find an overlap probability of 38\%. \par 
For higher momenta, increasing stiffness of the beam results in smaller separation to 
the TR photons and stronger rejection of detected clusters. As a consequence, the number of 
reconstructed TR clusters drops to 0.43 for 3~GeV/c.

\section{TR energy}

To relate the measured charge to the corresponding photon energy we compare the charge 
deposit in pion runs at B=0 to the simulated energy deposit \cite{dEdx}. To avoid any bias by 
single track space charge effects, which are maximal at perpendicular beam incidence, 
we use the charge collected at the beginning of the drift time, in the amplification 
region of the DC. The calibration factors obtained for each momentum 
from the most probable values (m.p.v.) of the measured charge and simulated energy spectra 
agree to an accuracy of 2.7\%. The main sources of errors are: 
1) the uncertainty of the assignment of the average pulse height distribution to the amplification region, 
which is determined for each run by variation of the interval of charge summation by $\pm$1 time bin. It is 
typically 10\%. 2) the deviation of the measured relative to the simulated shape of the charge spectrum, 
due to the inhomogeneous field in the amplification region and lack of statistics, resulting in an error 
of 2.5\% in determining the m.p.v. \par

\vspace{-.8cm}
\begin{figure}[ht]
\begin{tabular}{c} 
\begin{minipage}{.43\textwidth}
\includegraphics[width=1.0\textwidth]
{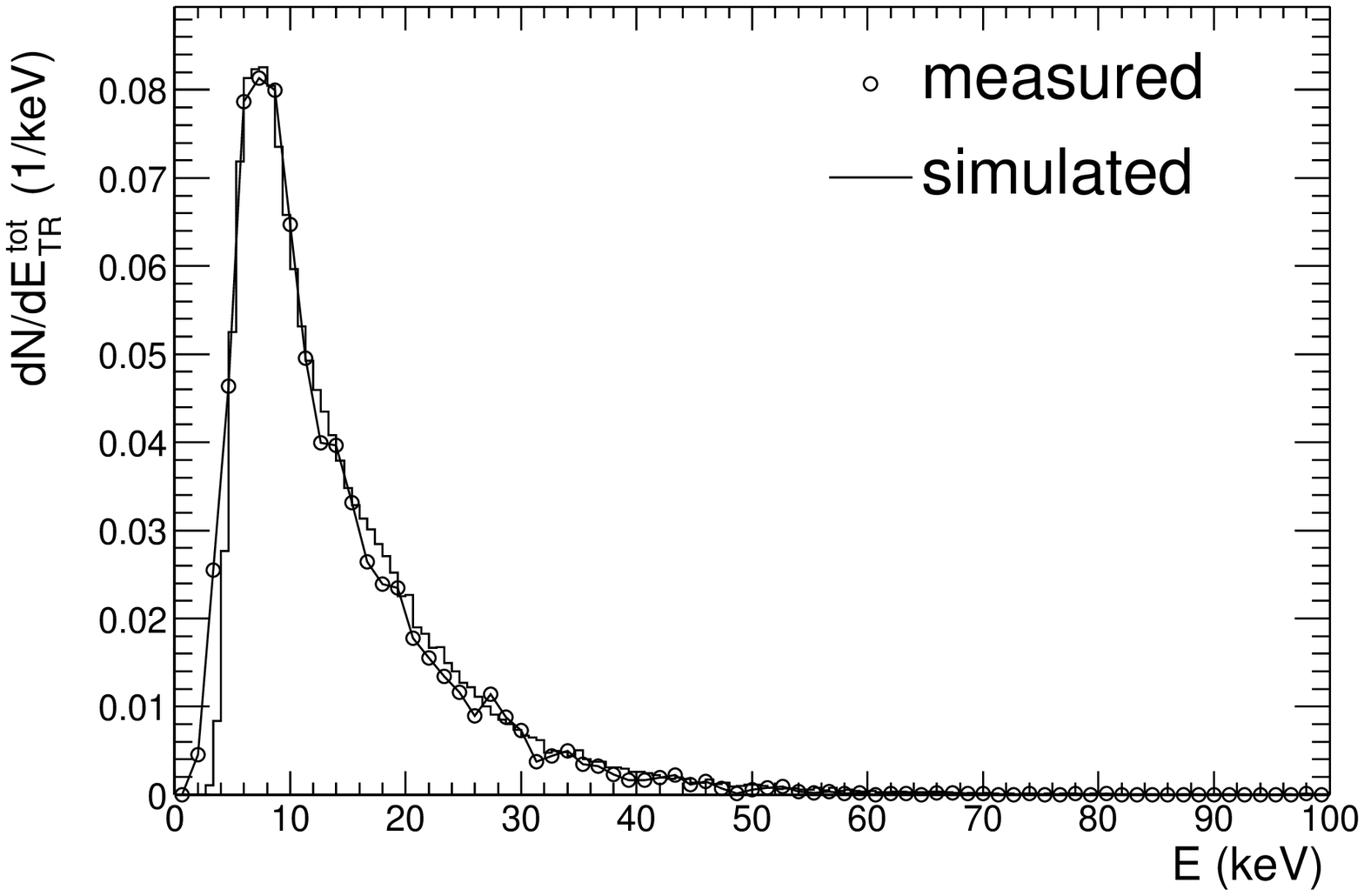}  
\end{minipage} \\
\begin{minipage}{.43\textwidth}
\includegraphics[width=1.0\textwidth,bbllx=0,bblly=0,
bburx=567,bbury=355,clip=]
{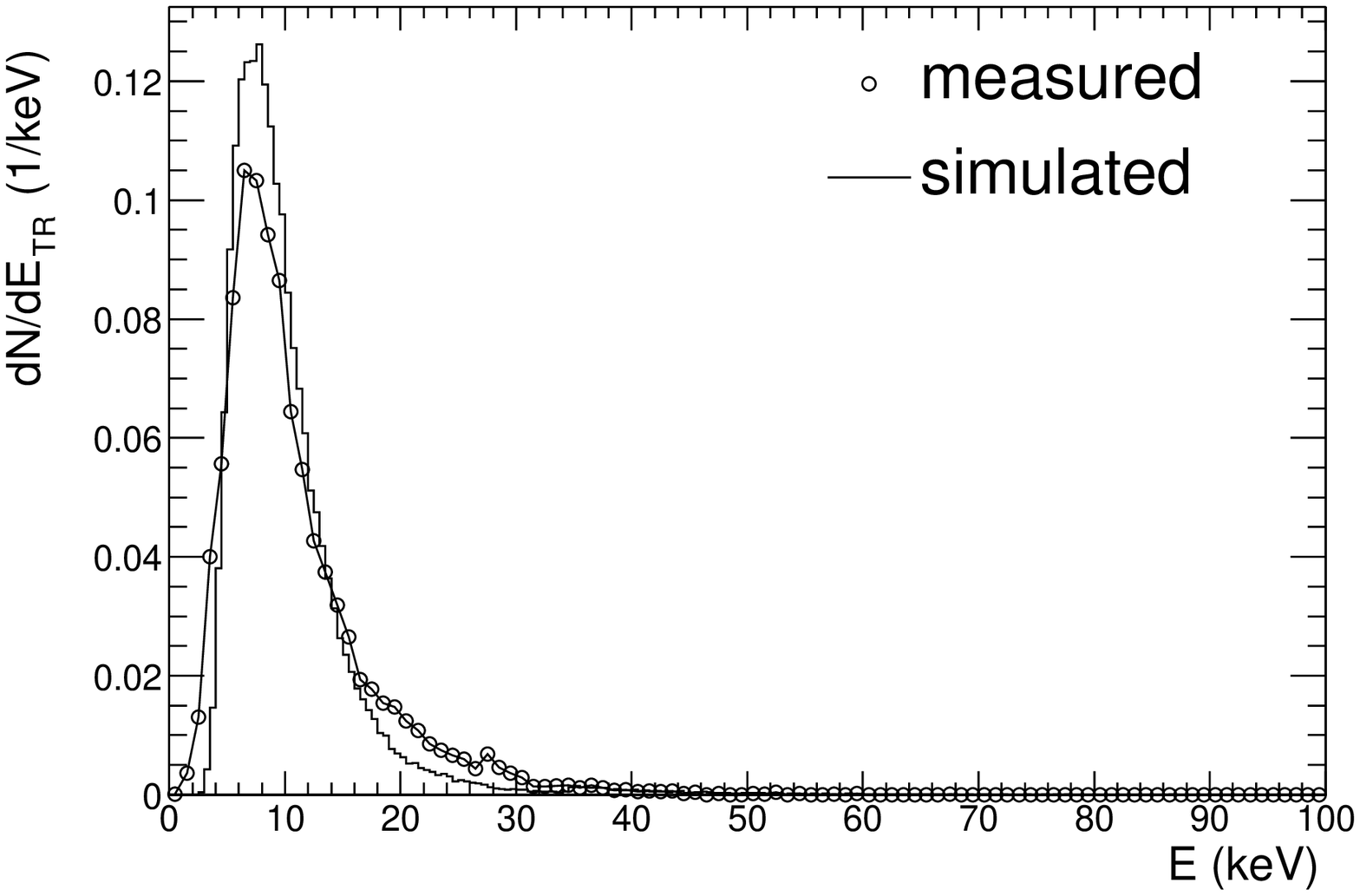} 
\end{minipage}
\end{tabular}
\vspace{-0.7cm}
\caption{Spectra of total TR energy (upper panel) and energy per photon (lower panel) 
for 2~GeV/c electron momentum.}
\label{TRspec}
\end{figure}

\vspace{-.5cm}

\begin{figure}[ht]
\begin{tabular}{c} 
\begin{minipage}{.43\textwidth}  
\vspace{-.8cm}\centering\includegraphics[width=1.0\textwidth, clip=]
{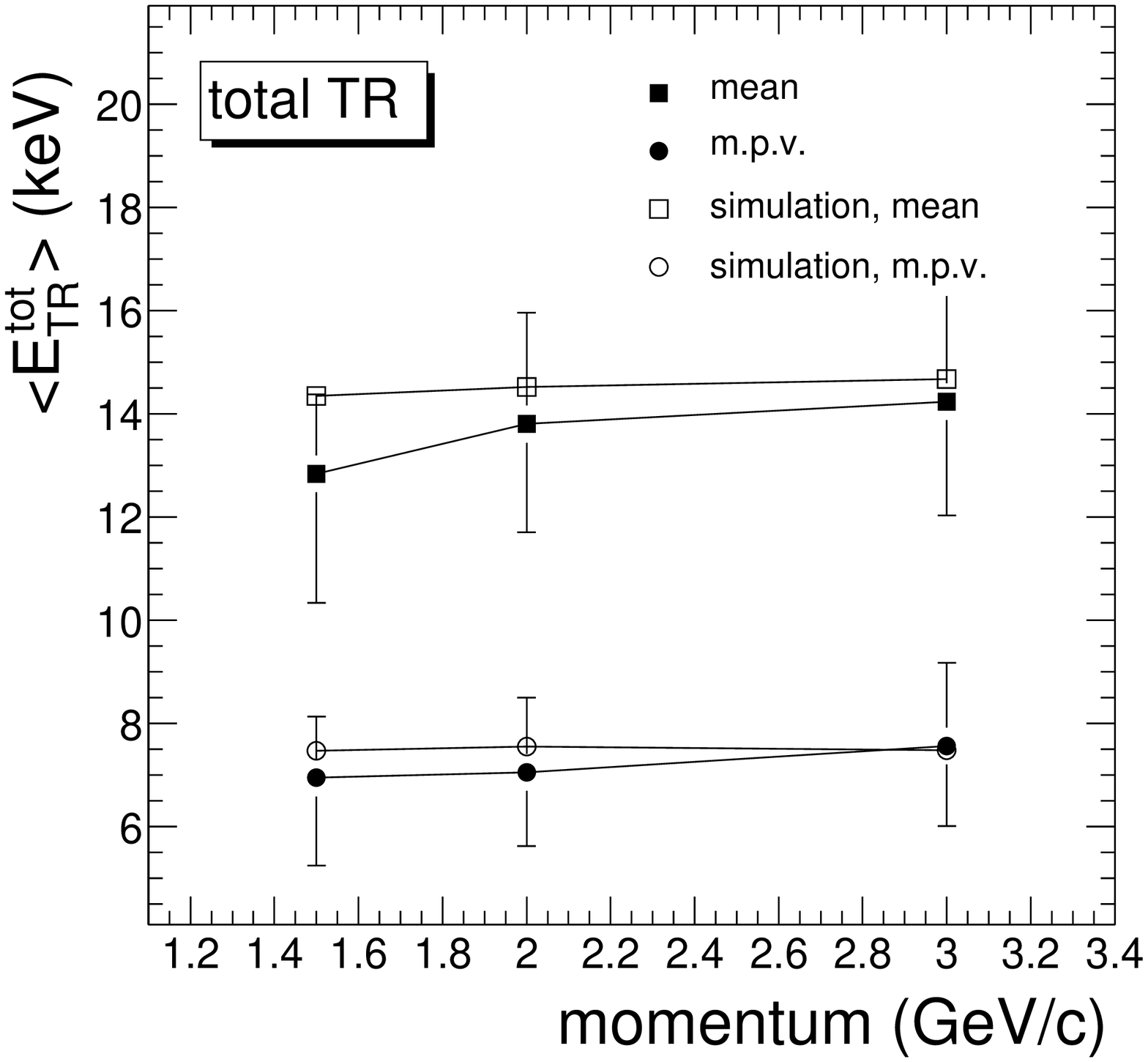} 
\end{minipage} \\
\begin{minipage}{.43\textwidth}
\centering\includegraphics[width=1.0\textwidth,clip=]
{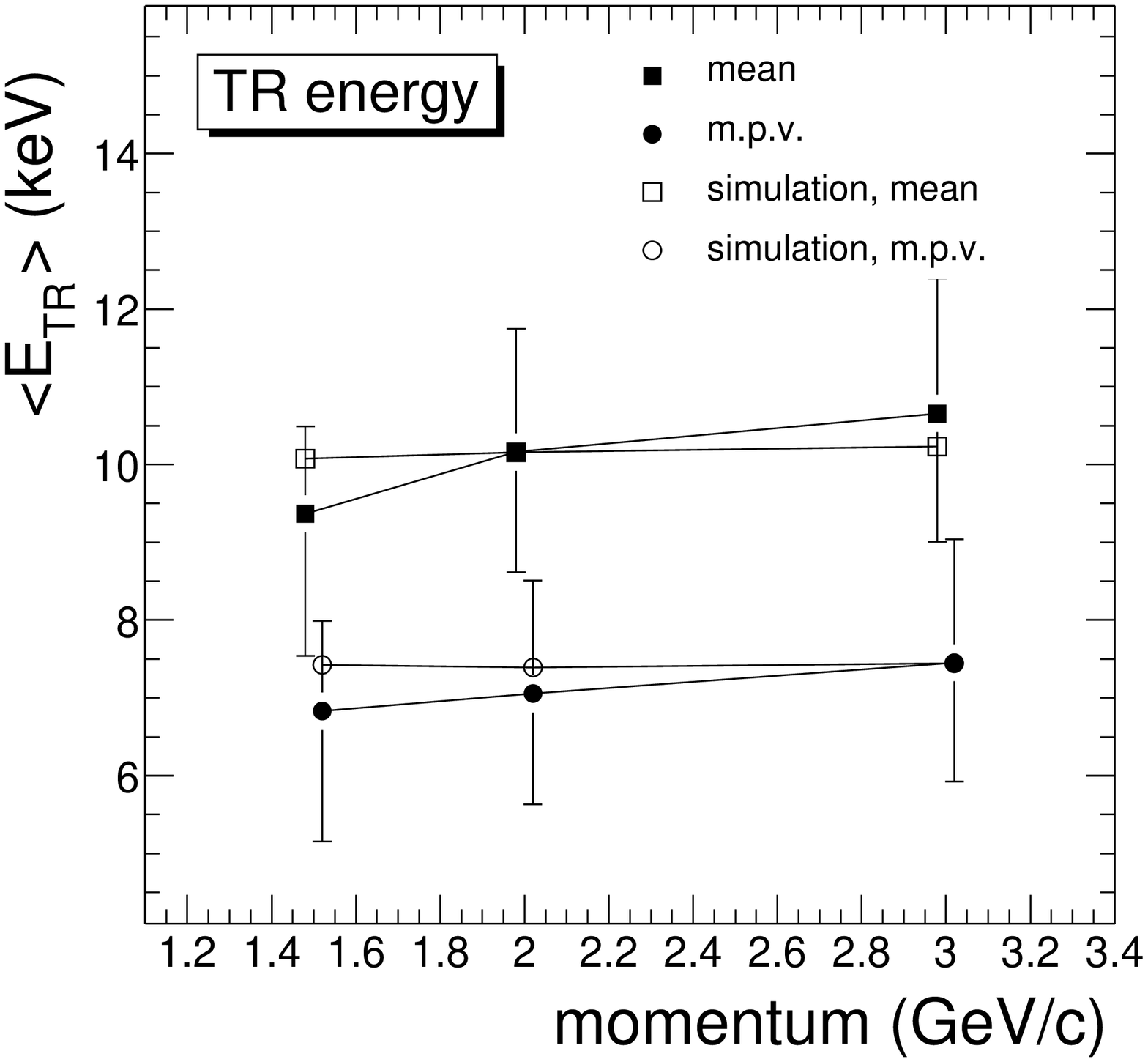} 
\end{minipage}
\end{tabular}
\vspace{-0.7cm}
\caption{Mean and most probable value of the spectra of total TR (upper panel) and 
single photon energy (lower panel). The data are compared to simulations.}
\label{ETR}
\end{figure}

In Fig.~\ref{TRspec} we present the spectra of total TR energy and 
energy per photon for a beam momentum of 2~GeV. 
The simulations reproduce the total TR spectrum, whereas the single photon 
spectrum has a more pronounced tail towards higher energies than 
calculated, as a consequence of cluster overlap. The evolution of the mean and m.p.v. of the 
spectra as function of momentum is shown in Fig.~\ref{ETR}. The errors on the data points are 
a 5\% uncertainty on the measured charge, reflecting the tolerance of the TR search 
algorithm to contamination from ionization, and the error of the energy calibration. 
The m.p.v. of the spectra is determined by a gaussian fit to the maximum. An additional 
error of 5\% on the m.p.v. accounts for the variation of the fit with the 
fit interval. Within the measurement errors, which are dominated by the systematic error of 
the calibration, the simulations agree with the measured values. 
Consistently with \cite{AntonProc} we observe a systematic increase 
of the TR yield as function of momentum. This trend is not 
reproduced by the simulations. 

\section*{ALICE TRD collaboration}
\vspace{-.1cm}
C.~Adler$^1$, A.~Andronic$^2$, V.~Angelov$^3$, H.~Appelsh{\"a}user$^2$, C.~Blume$^4$, 
P.~Braun-Munzinger$^2$, D.~Bucher$^5$, O.~Busch$^2$, V.~C{\u a}t{\u a}nescu$^6$, 
V.~Chepurnov$^7$, S.~Chernenko$^7$, M.~Ciobanu$^6$, H.~Daues$^2$, D.~Emschermann$^1$, 
O.~Fateev$^7$, P.~Foka$^2$, C.~Garabatos$^2$, R.~Glasow$^5$, T.~Gunji$^8$, M.~Gutfleisch$^3$, 
H.~Hamagaki$^8$, J.~Hehner$^2$, N.~Heine$^5$, N.~Herrmann$^1$, M.~Inuzuka$^8$, E.~Kislov$^7$, 
V.~Lindenstruth$^3$, C.~Lippmann$^2$, W.~Ludolphs$^1$, T.~Mahmoud$^1$, 
A.~Marin$^2$, D.~Miskowiec$^2$, K.~Oyama$^1$, Yu.~Panebratsev$^7$, V.~Petracek$^1$, 
M.~Petrovici$^6$, A.~Radu$^6$, C.~Reichling$^3$, K.~Reygers$^5$, A.~Sandoval$^2$, R.~Santo$^5$, 
R.~Schicker$^1$, R.~Schneider$^3$, K.~Schwarz$^2$, S.~Sedykh$^2$, R.S.~Simon$^2$, 
L.~Smykov$^7$, H.K.~Soltveit$^1$, J.~Stachel$^1$, H.~Stelzer$^2$, H.~Tilsner$^3$, G.~Tsiledakis$^2$, 
I.~Rusanov$^1$, W.~Verhoeven$^5$, B.~Vulpescu$^1$, J.~Wessels$^5$, B.~Windelband$^1$, 
V.~Yurevich$^7$, Yu.~Zanevsky$^7$ and O. Zaudtke$^5$
\par
\vspace{0.1cm}
$^1$Physikalisches Institut, Heidelberg, Germany; $^2$GSI, Darmstadt, Germany; 
$^3$Kirchhoff Institut, Heidelberg, Germany; $^4$Universit{\"a}t Frankfurt, Germany; 
$^5$Universit{\"a}t M{\"u}nster, Germany; $^6$NIPNE Bucharest, Romania; 
$^7$JINR Dubna, Russia; $^8$University of Tokyo, Japan.

\end{document}